\documentclass[lettersize,journal]{IEEEtran}
\usepackage{amsmath,amsfonts}
\usepackage{algorithm}
\usepackage{algpseudocode}  
\usepackage{array}
\usepackage[caption=false,font=normalsize,labelfont=sf,textfont=sf]{subfig}
\usepackage{textcomp}
\usepackage{stfloats}
\usepackage{url}
\usepackage{verbatim}
\usepackage{graphicx}
\usepackage{cite}
\usepackage{caption}  
\usepackage{enumitem}
\usepackage[acronym]{glossaries}
\usepackage{multirow}

\newacronym{mimo}{MIMO}{multiple-input multiple-output}
\newacronym{bs}{BS}{base station}
\newacronym{csi}{CSI}{channel state information}
\newacronym{ofdm}{OFDM}{orthogonal frequency division multiplexing}
\newacronym{ue}{UE}{user equipment}
\newacronym{dl}{DL}{deep learning}
\newacronym{csiadanet}{CsiAdaNet}{CSI Adaptive Network}
\newacronym{ai/ml}{AI/ML}{Artificial Intelligence/ Machine Learning}
\begin{document}
\title{Deep Learning assisted Port-Cycling based Channel Sounding for Precoder Estimation in Massive MIMO Arrays}

\author{\IEEEauthorblockN{Advaith Arun, Shiv Shankar, Dhivagar Baskaran, Klutto Milleth, Bhaskar Ramamurthi}\\
\IEEEauthorblockA{\textit{Centre of Excellence in Wireless Technology IIT Madras, India} \\
\{advaith22, shivshankar, dhivagar.b, klutto\}@cewit.org.in}}

\maketitle

\begin{abstract}
Future wireless systems are expected to employ a substantially larger number of transmit ports for \gls{csi} estimation compared to current specifications.
Although this improves spectral efficiency, it also increases the resource overhead to transmit reference signals across the time-frequency grid, ultimately reducing achievable data throughput.
In this work, we propose an \gls{dl}-based \gls{csi} reconstruction framework that serves as an enabler for reliable \gls{csi} acquisition in future 6G systems. 
The proposed solution involves designing a port-cycling mechanism that sequentially sounds different portions of \gls{csi} ports across time, thereby lowering the overhead while preserving channel observability. 
The proposed \gls{csiadanet} model exploits the resulting sparse measurements and captures both spatial and temporal correlations to accurately reconstruct the full-port \gls{csi}. 
The simulation results show that our method achieves overhead reduction while maintaining high \gls{csi} reconstruction accuracy.
\end{abstract}


\vspace{-0.7em}

\section{Introduction}
\IEEEPARstart{M}{assive} \gls{mimo} system is a crucial technology for wireless communication systems, offering benefits including spatial multiplexing, capacity gain, etc. 
These benefits rely on accurate \gls{csi}, whose acquisition incurs substantial overhead with increasing channel dimensions. 
In systems operating in frequency division duplexing (FDD) scenario, downlink channel acquisition is performed using \gls{csi}-RS \cite{1}, which enables reporting of channel quality, number of independent streams, and the beamforming vector \cite{2}.
However,  \gls{csi} resource overhead grows linearly with the array size, reducing data throughput. 

To address the growing overhead, industry and academic efforts have explored \gls{ai/ml} techniques for \gls{csi} feedback.
In particular, 3GPP has identified \gls{csi} compression and prediction as initial use-cases for \gls{ai/ml} based air interface \cite{3}.
In academia, CsiNet demonstrated effective non-linear \gls{csi} compression \cite{4}, while 
subsequent works investigate \gls{csi} extrapolation across time, frequency and spatial domains \cite{5}.
However these approaches primarily reduce feedback or temporal overhead and do not address the instantaneous overhead present due to large scale antenna sounding.

Motivated by this observation, this work focuses on reducing instantaneous channel sounding overhead in the antenna dimension, while maintaining compatibility with current standard compliant transmission. 
Unlike the environment based channel prediction approach that rely on site-specific information (i.e. images) \cite{6}, we adopt a port-cycling based channel sounding strategy in which a subset of ports are activated at a particular time instant, and the full panel is sequentially covered over multiple cycles. 

Building on this framework, we propose \gls{csiadanet}, a deep learning model that exploits the temporal and spatial correlations across the temporally aggregated \gls{csi} measurements to predict the quantized Type-II precoder defined in 5G NR \cite{1},\cite{7},enabling precoder estimation without requiring full-port instantaneous \gls{csi}. 

The main contributions of this work are summarized as follows:
\begin{enumerate}
    \item A port-cycling based \gls{csi} sounding technique that reduces the instantaneous spatial overhead in massive \gls{mimo} systems.
    \item A deep learning model that directly predicts Type-II precoder quantities from temporally aggregated port-cycled measurements.
    \item Link-level simulation in a 3GPP compliant setup demonstrate proposed approach achieves near optimal performance.
\end{enumerate}

The paper is organized as follows: The system model, port cycling scheme are introduced in Section-II. In Section-III, the details behind the \gls{csiadanet} model are described. The simulation setup and the results are described in Section-IV. The conclusions are summarized in Section-V.

\vspace{-0.9em}

\section{System Model}
Consider a downlink massive \gls{mimo} system with $N_t \gg 1$ transmit antennas with a uniform planar array at the \gls{bs} and $N_r$ receive antennas at the UE.
The system is operated in \gls{ofdm} with $K$ subcarriers. 
The received signal at the $k^{th}$ subcarrier can be expressed as 
\begin{equation}
 \mathbf{y}_{k} = \mathbf{h}_{k}\mathbf{v}^{H}_{k}\mathbf{x}_{k}+\mathbf{n}_{k},     
\end{equation}
where $\mathbf{y}_{k} \in \mathbb{C}^{N_{r} \times 1},\mathbf{h}_{k} \in \mathbb{C}^{N_r \times N_t},\mathbf{v}_k \in \mathbb{C}^{N_L \times N_t}, \mathbf{x}_{k} \in \mathbb{C}^{N_L \times 1}, n_{k}$ denotes the received signal, channel matrix, precoding vector, data vector and the additive noise of the $k^{th}$ subcarrier. 
The \gls{bs} is equipped with $N_t = 2 \times N_x \times N_y$ antennas, where $N_x,N_y$ denotes the number of dual polarized antenna elements at the horizontal and vertical dimensions.
The wideband covariance matrix is computed as 
\begin{equation}
\mathbf{R} = \frac{1}{K}\sum_{i=1}^{k}h^{H}_{i}h_i   ,
\quad \mathbf{R} \in \mathbb{C}^{N_t \times N_t}
\end{equation}
\subsection{Type II codebook in 5G NR}
5G-NR defines the configurable Type-II of codebooks based on a 2D-DFT based grid of beams in horizontal and vertical dimensions of the antenna array.  
 It selects the optimal $L$ orthogonal beams and linearly combines them in such a way that it maximizes the energy in the eigen vector direction. 
 The Rel.15 Type-II codebook \cite{1} for a given layer $\nu$ and the entire wideband is defined as: 
\begin{equation}
        \mathbf{W}^{(\nu)} 
        = \boldsymbol{\Lambda}^{(\nu)} 
        \left[
        \begin{array}{c}
        \displaystyle \sum_{i=0}^{L-1} {v}_{l^{(i)} m^{(i)}} 
        p_{\nu,i} \varphi_{\nu,i} \\[10pt]
        \displaystyle \sum_{i=0}^{L-1} {v}_{l^{(i)} m^{(i)}} 
        p_{\nu,i+L}\varphi_{\nu,i+L}
        \end{array}
        \right]
\end{equation}
\begin{table}[h!]
\centering
\begin{tabular}{|>{\centering\arraybackslash}p{0.4\columnwidth}|
                >{\centering\arraybackslash}p{0.4\columnwidth}|}
\hline
\textbf{Quantity} & \textbf{Description} \\ \hline
$v_{l,m}$ &
IDFT beam, $v_{l,m} = v_{l} \otimes u_{m}$ \\
&
$v_l$: horizontal beam; $u_m$: vertical beam \\ \hline
$L$ &
Number of beams, $L \in \{2,3,4\}$ \\ \hline
$p_{\nu}$ &
Wideband amplitude \\ \hline
$\varphi_{\nu}$ &
Wideband phase \\ \hline
\multirow{2}{*}{$\Lambda^{(\nu)}$} &
Normalization factor \\
&
$\Lambda^{(\nu)} = \frac{1}{\sqrt{N_x N_y \sum_{l=0}^{2L-1} (p_l)^2}}$ \\ \hline
\end{tabular}
\caption{Rel.15 Type-II quantities}
\end{table}

\begin{itemize}
\item \textbf{Beam Set}: It refers to the over sampled beam group out of the available $O_1O_2$ groups.
    \item \textbf{Beam Indices}: It refers to the top-$L$ orthogonal beams out of $N_1N_2$ beams within the selected beamset.
    \item \textbf{Wideband Amplitude}: It corresponds to reporting the quantized wideband amplitudes for the selected $2L-1$ orthogonal beams. The quantized amplitudes are drawn from the set, $P = {\{0,\sqrt{\frac{1}{64}},\sqrt{\frac{1}{32}},\sqrt{\frac{1}{16}},\sqrt{\frac{1}{8}},\sqrt{\frac{1}{4}},\sqrt{\frac{1}{2}},1\}}$     
    \item \textbf{Wideband Phase}:  This corresponds to the reporting of wideband phase of the selected $2L-1$ orthogonal beams. The strongest beam within the beam set is normalized to zero phase. The quantized phase alphabets are drawn from the set $\Phi = \{ e^{\frac{j2\pi n}{N_{psk}}}, n \in \{0,1,...,N_{psk}-1\}\}$
\end{itemize}

The beams are selected from within a beamset based on Algorithm~\ref{alg:beam_set_ind_sel} \cite{8} and the wideband components are computed based on Algorithm~\ref{alg:wb_amp_ph_selection}.

\begin{algorithm}
\caption{Beam Set and Beam Indices Selection}
\label{alg:beam_set_ind_sel}
\begin{algorithmic}[1]
\State \textbf{Input:} $\{\mathbf{V}_b\}_{b=1}^{B}$, $\mathbf{R}$, $\mathbf{L}$
\State \textbf{Output:} $B^\star, \mathcal{I}^\star$
\For{$b=1$ to $B$}
    \State $\mathbf{r}\!\gets\!\mathbf{V}_b^{H}(\mathbf{R}_{11}+\mathbf{R}_{22})\mathbf{V}_b$
    \State $\mathcal{I}_b\!\gets\!\operatorname{Top}L(|\operatorname{diag}({r})|)$
    \State $\eta_b\!\gets\!\sum_{i\in\mathcal{I}_b}|{r}_{i}|$
\EndFor
\State $B^\star\!\gets\!\arg\max_b\eta_b,\ \mathcal{I}^\star\!\gets\!\mathcal{I}_{B^\star}$
\State \Return $B^\star,\mathcal{I}^\star$
\end{algorithmic}
\end{algorithm}

\begin{algorithm}
\caption{Wideband Amplitude and Phase Selection}
\label{alg:wb_amp_ph_selection}
\begin{algorithmic}[1]
\State \textbf{Input:} $b^\star,\mathcal{I}^\star,\mathbf{R}$ \quad
\textbf{Output:} $\mathcal{P}^\star_\nu,{\varphi}^\star_\nu$
\State $({\Lambda},{E})\!\gets\!\mathrm{EVD}(\mathbf{R})$,
$\mathbf{e}_{0,\nu}\!=\!\mathbf{E}[1{:}N_t/2,\nu]$,
$\mathbf{e}_{1,\nu}\!=\!\mathbf{E}[N_t/2{+}1{:}N_t,\nu]$
\For{$l=1$ to $L$}
    \State $p^\nu_{0,l}\!=\!\mathbf{e}^H_{0,\nu}\mathbf{v}_l,\ 
           p^\nu_{1,l}\!=\!\mathbf{e}^H_{1,\nu}\mathbf{v}_l$
\EndFor
\For{$\text{pol}\!\in\!\{0,1\}$}
    \State $\tilde{\mathbf{a}}^\nu_{pol}\!\gets\!
    |p^\nu_{pol}|/\max_k |p^\nu_{k,pol}|$
    \State ${\theta}_{\nu,pol}\!\gets\!\angle\tilde{\mathbf{a}}^\nu_{pol}$
    \State $\mathcal{A}^\nu_{pol}\!\gets\!\arg\min_i|P_i-\tilde{\mathbf{a}}^\nu_{pol}|$
    \State ${\Theta}^\nu_{pol}\!\gets\!\arg\min_j|\Phi_j-{\theta}_{\nu,pol}|$
\EndFor
\State $\mathcal{P}^\star_\nu\!\gets\!P[\mathcal{A}^\nu],\ 
       \boldsymbol{\varphi}^\star_\nu\!\gets\!\Phi[{\Theta}^\nu]$
\State \Return $\mathcal{P}^\star_\nu,{\varphi}^\star_\nu$
\end{algorithmic}
\end{algorithm}

\vspace{-0.3em}
\subsection{Sub-array partitioning}
To reduce the number of active ports for sounding, the uniform planar array (UPA) of size ($N_x \times N_y$) is partitioned into $\rho = \rho_x \times \rho_y$ non-overlapping sub-panels.
Each sub-panel consists of $N'_x = N_x/\rho_x$ and $N'_y = N_y/\rho_y$, resulting in $N'_t = N'_x\times N'_y \times N_{pol}$ ports per sub-array.

\begin{figure}[H]
\centering
\includegraphics[width=0.8\columnwidth]{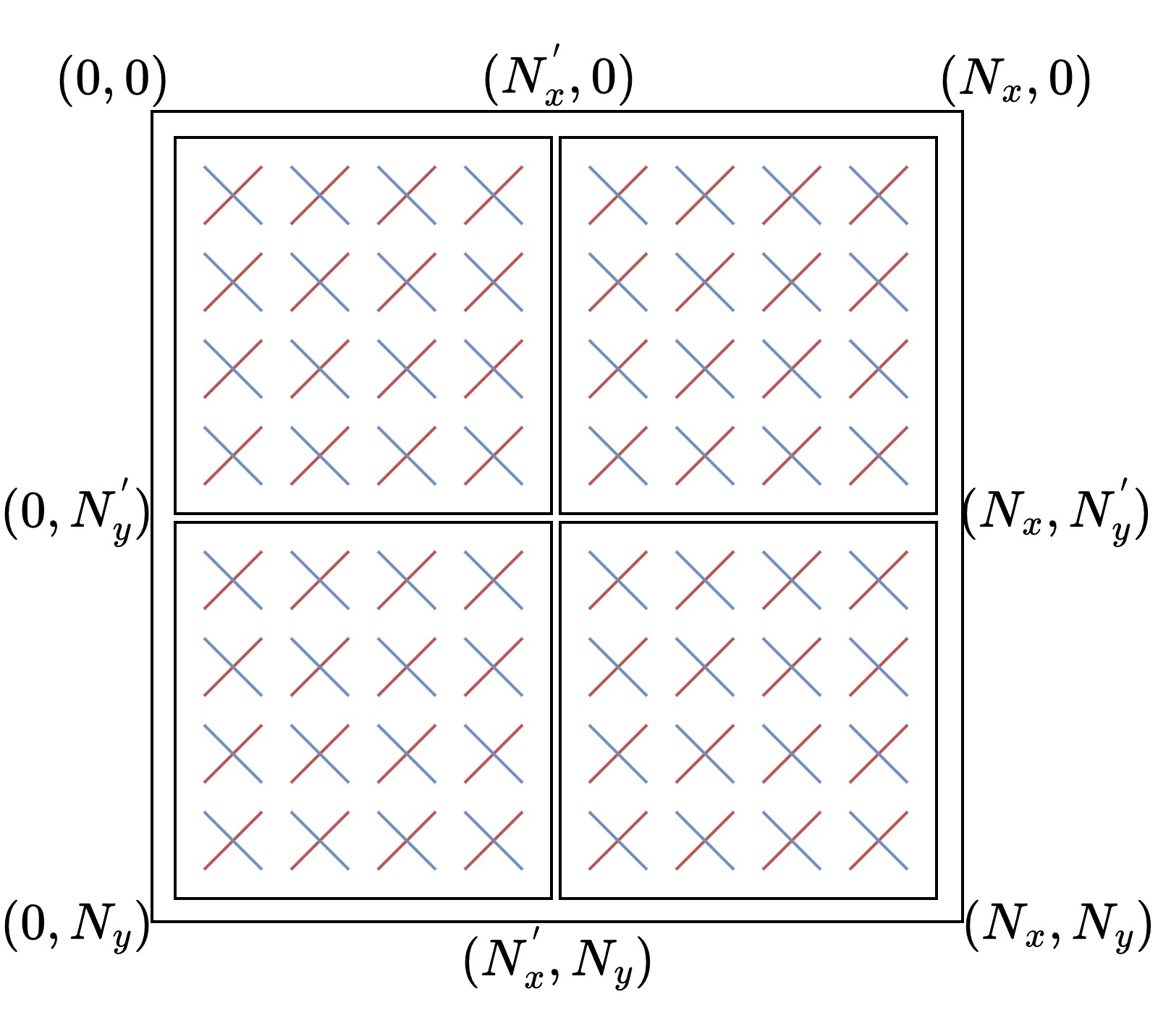}
\caption{Illustration of sub-panel based partitioning}
\label{fig:subpanel_partitioning}
\end{figure}

Sub-panel based partitioning, preserves the original antenna spacing, and thereby avoids spatial aliasing in the beam computation, making it well suited for port-cycling based \gls{csi} acquisition. An illustrative example of the procedure is shown in Fig.~\ref{fig:subpanel_partitioning}.

\vspace{-1.3em}
\subsection{Port Cycling Scheme}
It refers to the sequential activation of non-overlapping sub-arrays across multiple time instances for downlink \gls{csi} transmission.
At each instant, only one sub-array is sounded and the full antenna array is covered after $\rho$ cycles. 
This enables instantaneous reduction in spatial overhead while retaining the observability of the full array across time.

The feasibility of this method relies on the temporal coherence of beam related quantities during the port-cycling duration. To quantify validity of this scheme, a \textit{variation score} ($\mathcal{M}$) is defined as
\begin{figure}[H]
\centering
\includegraphics[width=0.8\columnwidth]{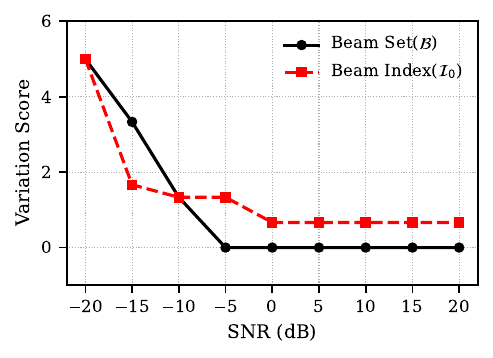}
\caption{Plot of variation scores}
\label{fig:var_scores}
\end{figure}
\vspace{-0.2em}
    \[
    \text{Variation Score} 
    = \frac{1}{\rho - 1} \sum_{i=1}^{\rho-1} \left| \mathcal{M}^{\,i+1} - \mathcal{M}^{\,i} \right|.
\]
\vspace{-0.5em}

Fig.~\ref{fig:var_scores}, illustrates that both beamset and dominant beam index exhibit low variation across port-cycling instances, particularly at moderate-to-high SNR region, validating the feasibility of temporally aggregating port-cycled CSI measurements to predict precoder.

\section{\gls{csiadanet}}
This section presents a deep learning assisted solution for port cycling based system, where 
the proposed CSIAdaNet resides at the \gls{ue} side and operates on $\rho$ periodic \gls{csi} sub-panel measurements, to predict the quantized Type-II codebook parameters.

At a given prediction instance, the operation of \gls{csiadanet} can be expressed as
\begin{equation}
 \{B^\star,\; \mathcal{I}^\star,\; \mathcal{P}^\star, \mathcal{\varphi}^\star\} = \mathcal{F}(\{\mathcal{H}_{t+iP}\}_{i=0}^{\rho-1};\theta) 
\end{equation}
where $\mathcal{F}(\cdot)$ denotes the \gls{dl} model, with parameters $\theta$  and $\mathcal{H}_{t+iP}$ denotes the \gls{csi} measurements corresponding to sub-array sounded at time $t+iP$. 

The model performs the predictions hierarchically with contextual dependency. 
The beamset is first predicted, followed by the top $L$ beam indices conditioned on the beamset, and finally the quantized wideband amplitude and phase levels.

The input to the model consists of $\rho$ sub-array CSI tensors, $\mathcal{H}_{t} \in \mathbb{R}^{N_x' \times N_y' \times (N_rN_{sc}) \times (2N_{pol})}$, where the factor of $2$ corresponds to the real and imaginary components of the complex channel co-efficients.

\begin{figure*}[t]
    \centering
    \includegraphics[width=\textwidth]{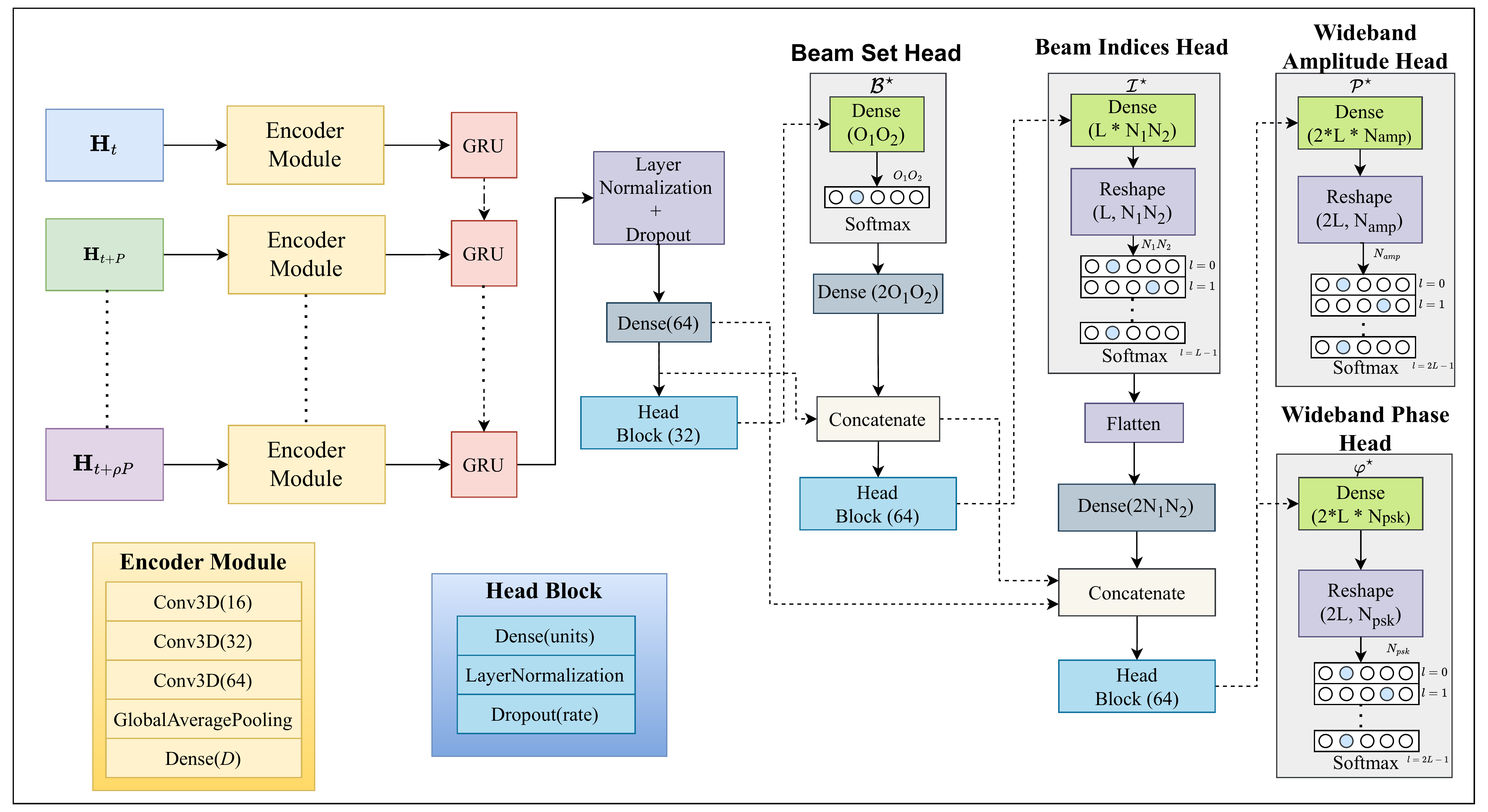}
    \captionsetup{justification=centering}
    \caption{Model Architecture of CSIAdaNet}
    \label{fig:model_arch}
\end{figure*}

Each CSI tensor is processed by a shared encoder module $\mathrm{Enc}(\cdot)$ to extract the spatial-frequency features while preserving the sub-array structure $(N'_x \times N'_y)$.
It comprises of three 3D convolutional layers with filter sizes \{16,32,64\} followed by a global average pooling and a dense layer of $D= 32$ units, yielding a latent representation $\mathbf{z}_t = \mathrm{Enc}(\mathcal{H}_t)\in \mathbb{R}^{D}$.
The sequence $\{\mathbf{z}_t\}_{t=1}^{\rho}$ is further processed by a Gated Recurrent Unit (GRU) to capture temporal correlations across port-cycling instances and further feature refinement through Head Block units. 
The aggregated representation is then fed to multiple task-specific heads, each implemented as a feed-forward classifier operating on the shared latent features.The overall model architecture of \gls{csiadanet} is illustrated in Fig.~\ref{fig:model_arch}.

The \textit{Beam Set Head} predicts the over-sampled beamset index ($\mathcal{B}$), providing a coarse spatial representation . 
The head operates on the shared representation and outputs a probability distribution over the $O_1 \times O_2$ beamset levels. 
Conditioned on this, the \textit{Beam Indices Head} predicts the top $L$ beam indices ($\mathcal{I}$) within the selected beamset.
The \textit{Wideband Amplitude Head} predicts the quantized wideband amplitude levels for the selected $2L$ beams from a pre-defined set $P$, while the \textit{Wideband Phase Head} predicts the quantized phase levels from $\Phi$ for coherent beam combination. 

The CSIAdaNet is trained using a weighted multi-task sparse categorical cross-Entropy (SCCE) loss. 
For each \gls{csi} quantity ($x \in \{ \mathcal{B},\mathcal{I},\mathcal{P},\mathcal{\varphi}\} $), the task specific loss is defined as $ \mathcal{L}_x = \mathbb{E}[\mathit{l}_{SCCE}(\hat{x},\mathit{p}_x)]$,
where $\hat{x}$ denotes the ground-truth class index and $\mathit{p}_x$ is the predicted class probability score.
The overall training loss is defined as 
\begin{equation}
\mathcal{L}_{total} = \sum_{x \in \{\mathcal{B},\mathcal{I},\mathcal{P},\mathcal{\varphi}\}}\lambda_x\mathcal{L}_x
\end{equation}
Finally, the predicted codebook can be reconstructed as 
\begin{equation}
\mathbf{\tilde{W}}^{(\nu)} 
= \boldsymbol{\Lambda}^{(\nu)} 
\left[
\begin{array}{c}
\displaystyle \sum_{i=0}^{L-1} {V_b[\mathcal{B}^{\star}][\mathcal{I}^{\star}_{i}]} \times 
\mathcal{P}^{\star}_{\nu,i} \times  e^{j\mathcal{\varphi}^{\star}_{\nu,i}} \\[8pt]
\displaystyle \sum_{i=0}^{L-1} V_b[\mathcal{B}^{\star}][\mathcal{I}^{\star}_{i}] \times
\mathcal{P}^{\star}_{\nu,i+L} \times e^{j\mathcal{\varphi}^{\star}_{\nu,i+L}}
\end{array}
\right]
\end{equation}
\vspace{-0.6em}

\section{Simulation and results}
In this section, we discuss the simulation framework considered. We use a 3GPP standards compliant proprietary simulator based on 5G NR Rel.19. 
\vspace{-0.5em}
\subsection{Link-Level Simulation Assumptions}

The simulation is conducted using a 3GPP compliant link-level simulator under an Urban Macro (UMa) line-of-sight scenario with an 200m inter-site distance \cite{9}. 
A single-cell, three-sector deployment is considered, with a 128-antenna uniform planar array at the BS and four antennas at the UE.
The BS array is partitioned into sub-panels with spatial reduction factor ($\rho=4$) and the CSI-RS periodicity is set to 20 ms, and single layer transmission is assumed, i.e  ($\nu = 1$).
The key simulation parameters are summarized in Table II.

\begin{table}[H]
\begin{tabular}{|c|c|}
\hline
\textbf{Parameter} & \textbf{Value} \\ \hline
\text{Deployment Scenario} & Urban Macro (UMa) \textit{LoS} \\ \hline
\text{Number of transmit antennas} ($N_t$) & $128$ \\ \hline 
\text{Transmit antenna array configuration} & $[N_x=8, \;N_y=8, \;N_{pol}=2]$ \\ \hline
\text{Number of receive antennas} ($N_r$) & $4$ \\ \hline
\text{\gls{csi}-RS periodicity} ($T_{{csi}}$) & $20\; \text{ms}$ \\ \hline
\text{Carrier frequency} ($f_c$) & $3.5 \;\text{GHz}$ \\ \hline
\text{System bandwidth} (BW)& $10 \;\text{MHz}$ \\ \hline
\text{Subcarrier spacing} ($scs$)& $15\; \text{kHz}$ \\ \hline
\text{UE velocity} & $30 \;km/h$ \\ \hline
\text{Spatial reduction factor} ($\rho$) & $4~(\rho_x = 2,\; \rho_y = 2)$ \\ \hline
\text{Number of antennas per sub-array} ($N_t^{'}$) & 32 \\ 
\hline
\end{tabular}
\captionof{table}{Simulation Assumptions}
\end{table}

\vspace{-0.8em}
\subsection{\gls{csiadanet} Model}

The dataset consists of $30,000$ samples, split into training, validation, and test sets in a $70:20:10$ ratio.
To improve robustness, training is performed under multiple SNR conditions uniformly sampled from $\{-5, 0, 5, 10, 15\}$ dB, while evaluation is conducted over a wider SNR range of $-20$ to $20$ dB.
The model is trained for $250$ epochs using Adam optimizer with a learning rate of $10^{-3}$
and a batch size of $256$ employing a weighted multi-task loss, with a loss weights ($\lambda_{\mathcal{B}}: 0.75,\lambda_{\mathcal{I}}: 1.5, \;\lambda_{\mathcal{P}}: 0.75,\;\lambda_{\mathcal{\varphi}}:1.25)$.
A higher weightage is assigned to beam indices ($\lambda_{\mathcal{I}}$) to emphasize any error in that propagates directly in precoder construction.
All heads use sparse categorical cross-entropy with accuracy as the evaluation metric, while early stopping and adaptive learning-rate reduction are applied to prevent overfitting.
Sub-arrays are sequentially sounded for port cycling, with all ($\rho!$) sounding permutations used during training to ensure model robustness.
\vspace{-1em}
\subsection{Metrics}
As a traditional method, we consider the actual Type-II precoder obtained using the full set of ports ($W_{T2}$) as the reference for comparison, with the 
precoder quantities using Algorithms \ref{alg:beam_set_ind_sel} and \ref{alg:wb_amp_ph_selection}.

 \textbf{Squared Generalized Cosine Similarity}(SGCS): measures the directional alignment between the traditional or \gls{csiadanet} predicted Type-II precoder and the dominant eigen vector of the channel, with values ranging from 0 to 1. 
\begin{equation}
\mathrm{SGCS}(\mathbf{v}_{\text{eig}}, \mathbf{w}) = \frac{\left| \mathbf{v}_{\text{eig}}^{\mathrm{H}} \mathbf{w} \right|^2} {\lVert \mathbf{v}_{\text{eig}} \rVert_2^2 \, \lVert \mathbf{w} \rVert_2^2}, \mathbf{w} \in [\ \tilde{W},W_{T2}]\ 
\end{equation}
\textbf{Beamforming Gain} (BF): measures the effective power captured by the traditional or \gls{csiadanet} based Type-II precoder, relative to the optimal eigen-beamformer. 
\begin{equation}
Gain(\mathbf{v}_{ \mathbf{w},\text{eig}}) = \frac{\mathbf{w}^{H}\mathbf{R}\;\mathbf{w}}
     {\mathbf{v}_{\mathrm{eig}}^{H}\mathbf{R}\;\mathbf{v}_{\mathrm{eig}} \,}, \mathbf{w} \in [\ \tilde{W},W_{T2}]\ 
\end{equation}

\begin{figure*}[htbp]
    \centering
    \begin{minipage}{0.333\textwidth}
        \centering
        \includegraphics[width=\textwidth]{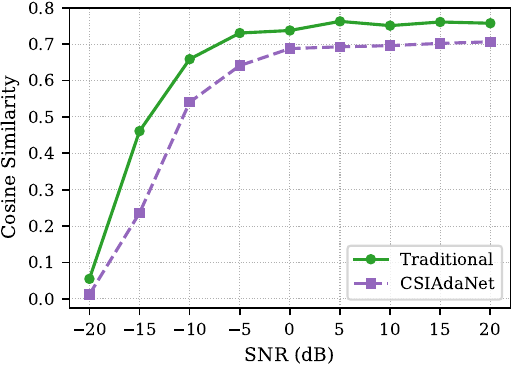}
        \caption*{(a) Cosine Similarity}
    \end{minipage}\hfill
    \begin{minipage}{0.333\textwidth}
        \centering
        \includegraphics[width=\textwidth]{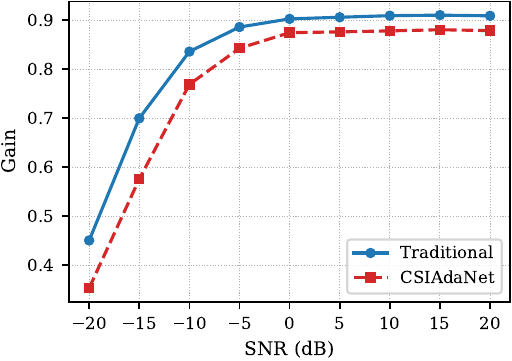}
        \caption*{(b) Beamforming Gain}
    \end{minipage}\hfill
    \begin{minipage}{0.333\textwidth}
        \centering 
        \includegraphics[width=\textwidth]{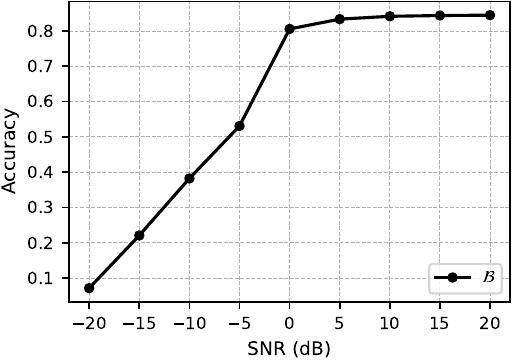}
        \caption*{(c) Beamset Accuracy}
    \end{minipage}

    \vspace{0.3cm}

    \begin{minipage}{0.333\textwidth}
        \centering
        \includegraphics[width=\textwidth]{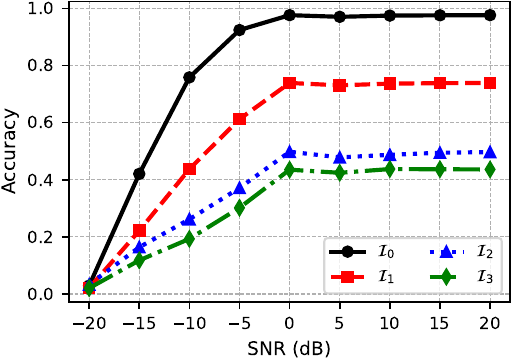}
        \caption*{(d) Beam Indices Accuracy}
    \end{minipage}\hfill
    \begin{minipage}{0.333\textwidth}
        \centering
        \includegraphics[width=\textwidth]{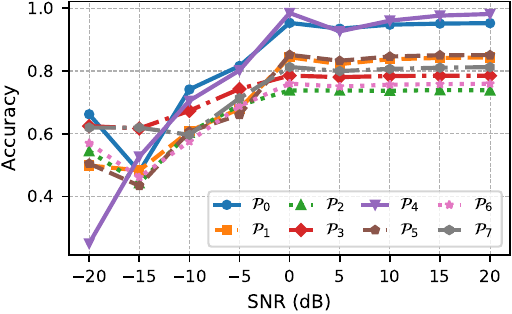}
        \caption*{(e) Wideband Amplitude Accuracy}
    \end{minipage}\hfill
    \begin{minipage}{0.333\textwidth}
        \centering
        \includegraphics[width=\textwidth]{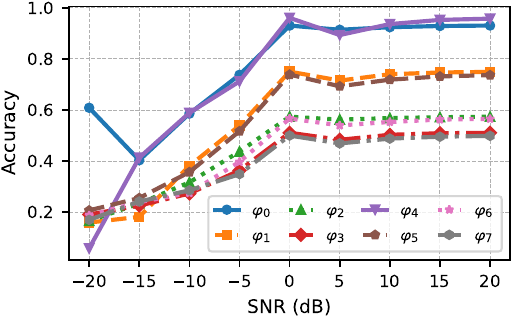}
        \caption*{(f) Wideband Phase Accuracy}
    \end{minipage}

    \caption{Performance evaluation of CSI Type-II precoder reconstruction across SNR: (a) cosine similarity, (b) beamforming gain, (c) beamset accuracy, (d) beam indices accuracy, (e) wideband amplitude accuracy, and (f) wideband phase accuracy. }
    \label{fig:performance evaluation}
\end{figure*}
where $\mathbf{v}_{eig}$ denotes the dominant eigen vector, $\tilde{W}$ refers to the \gls{csiadanet} predicted precoder from port-cycled channel measurements and $W_{T2}$ refers to the actual Type-II precoder corresponding to the actual number of ports. 
\vspace{-0.7em}
\subsection{Performance Evaluation and Analysis}
\subsubsection{Precoder Alignment and Beamforming Performance}
Figs. 5(a) and 5(b) evaluate the alignment and beamforming performance of \gls{csiadanet}. 
As shown in Fig. 5(a), the \gls{dl} based approach closely follows the conventional baseline across all SNRs, with SGCS increasing from below $0.2$ at low SNR to approximately $0.7$ in the medium-to-high SNR region.
Consistently, Fig. 5(b) shows that beamforming gain achieved using \gls{csiadanet}, exhibiting close matches to the traditional baseline, with a small gap at high SNR (within $\approx 0.03$).
The drop at low SNR is primarily due to noise-sensitive estimation.

\subsubsection{Analysis of CSI feedback quantities}
Figs. 5(c)–5(f) summarize the prediction accuracy of the individual \gls{csi} feedback quantities. 
In Fig. 5(c), beamset prediction remains highly reliable even at moderate SNR levels, due to its coarse spatial resolution and inherent robustness to noise, validating the port-cycling assumption.
In Fig. 5(d), the beam indices accuracy improves more gradually with SNR, reflecting the finer angular sensitivity. 
The wideband amplitude accuracy in Fig. 5(e) follows a similar trend, as amplitude estimation depends on the correctness of the beam selection.
Finally, Fig. 5(f) indicates that wideband phase prediction is more challenging at low SNR but achieves high accuracy at moderate-to-high SNR, demonstrating the effectiveness of temporally aggregated port-cycled CSI.
\subsubsection{Complexity Analysis}
This subsection compares the computational complexity of the \gls{csiadanet} model with conventional methods.
Conventional approaches require eigenvalue decomposition (EVD) of the full covariance matrix  $R \in \mathbb{C}^{N_t \times N_t}$, incurring $O(N^3_t)$ complexity.
In contrast, \gls{csiadanet} performs inference on sub-panel observations, with complexity $O(\rho D^2)$, independent of the full array size $N_t$, providing improved scalability for large-array systems.

\vspace{-0.5em}

\section{Conclusion}
This work presents a \gls{dl}-assisted port-cycling based precoder estimation that exploits both spatial diversity across antennas and temporal diversity arising  from port-cycled measurements to reduce instantaneous \gls{csi}  overhead in large MIMO systems. 
Rather than sounding all antenna ports simultaneously, spatially disjoint sub-panels are activated sequentially, and the partial \gls{csi} measurements are temporally aggregated .
The proposed \gls{csiadanet} model leverages these spatial–temporal correlations to reconstruct Type-II codebook quantities from reduced \gls{csi} observations.
Link-level simulations in a 3GPP-compliant setup demonstrate near-optimal beam alignment and beamforming gain at moderate to high SNR levels, with only marginal degradation under noisy regions.
Although all antenna ports are eventually sounded across multiple occasions, the number of active ports at any given instant is reduced by a factor of $\rho$, reducing instantaneous overhead and RF chain usage, highlighting a key design consideration for scalable and energy-efficient \gls{csi} acquisition in next-generation large-scale MIMO systems.

\vspace{-0.7em}
\bibliographystyle{IEEEtran}
\bibliography{IEEEabrv, refrence}

\end{document}